\documentclass[12pt]{article}
\usepackage{hyperref}
\usepackage[english]{babel}
\usepackage{amsmath}
\usepackage{latexsym}
\usepackage{amssymb}
\usepackage[all]{xy}
\usepackage[dvips]{graphicx}
\usepackage{epsfig}
\usepackage{psfrag}

\numberwithin{equation}{section}
\numberwithin{table}{section}
\numberwithin{figure}{section}

\begin{document}

\begin{titlepage}

\begin{center}

\hfill  hep-th/0807.2246  \\
\hfill ITFA-2008-23

\vskip 0.5 cm {\Large \bf Entropy Functions with 5D Chern-Simons terms  }

\vskip 1.25 cm {   Xerxes D. Arsiwalla  \footnote{e-mail:  X.D.Arsiwalla@uva.nl  }    }\\
{\vskip 0.5cm  Institute for Theoretical Physics\\ University of Amsterdam\\
Valckenierstraat 65\\
1018 XE, Amsterdam\\
The Netherlands\\}

\end{center}

\vskip 2 cm

\begin{abstract}
\baselineskip=17pt
In this note we reconsider Sen's entropy function analysis for 5D supergravity actions containing Chern-Simons terms. The apparent lack of gauge invariance is usually tackled via a 4D reduction. Here we motivate how a systematic 5D procedure also works. In doing so,  it becomes important to identify the correct 5D charges.  In particular, we perform explicit calculations for the black ring and 5D   black hole. In the black ring analysis, we find  Chern-Simons induced spectral flow shifts emerging out of Sen's formalism. We find that the entropy function nevertheless  remains  gauge invariant and  the resulting electric charges are identified as  Page charges.  For the  black hole too, 5D gauge invariance is  confirmed. Our 5D analysis enables us to fix  a mismatch that arose in the  electric charges of Goldstein and Jena's 4D-reduced calculation. Finally we provide an interpretation for the $e^0 \, \leftrightarrow \, p^0$ exchange in the entropy   function as  an interpolation between black hole and black ring geometries in Taub-NUT.  
\end{abstract} 
\end{titlepage}

\pagestyle{plain}
\baselineskip=19pt

\setcounter{page}{2}

\tableofcontents

\section{Introduction}         
The entropy function formalism of Sen \cite{S1}, \cite{S2}  allows for a very systematic approach to computing black hole entropy in $D$ dimensions with $AdS_2 \times S^{D - 2}$ near-horizon geometry, especially including higher derivative corrections.   Subsequently this formalism has also found application to other   extremal   black objects such as black rings and even  black holes with reduced near-horizon isometry groups   \cite{GJ}, \cite{CP}.  However, in odd dimensions, the presence of Chern-Simons terms in the supergravity action no longer leaves the latter invariant under large gauge transformations; whereas Sen's original construction   was formulated for gauge as well as reparametrization invariant  actions.  To overcome this hurdle, it was proposed in  \cite{SS1}  to perform  a dimensional reduction in order to bring the Lagrangian density  into a gauge invariant form and then apply the entropy function method.  Therefore whilst computing the black ring entropy function, the authors of  \cite{GJ}   first perform a  dimensional reduction of the 5D supergravity Lagrangian into a gauge invariant 4D Lagrangian, upon which the standard entropy function method can then be  applied. 

In this note we revisit the black ring and 5D static black hole entropy functions.  Instead of taking recourse to a  dimensional reduction, we propose that  a meaningful 5D computation of the entropy function with Chern-Simons terms is possible\footnote{ In this note we only consider gauge-type  Chern-Simons terms. Presumably our considerations are also valid for gravitational or mixed gauge-gravitational Chern-Simons terms.    }.   While  performing  such a 5D analysis, a key issue which requires careful consideration is how we should treat charges in 5D and their corresponding spectral flows. For the benefit of our esteemed reader, let us recall that these are also the same questions that have been at the center of much debate \cite{BK3},  \cite{BKW}, \cite{HR}, \cite{GHLS}, \cite{BK2} with regards to the 4D/5D conjecture for black holes and black rings \cite{GSY1}, \cite{GSY2}.  It is not surprising that those subtleties also come into play  when trying to perform an intrinsic 5D analysis of the entropy function formalism.  And that happens because the introduction of  Chern-Simons terms brings in three different notions of charge : Brane-source charge, Maxwell charge and Page charge \cite{DM}.  Which one is more relevant depends very much on the details of the geometric configuration one is in. Expressing the entropy function in terms of the correct 5D charges will turn out to be a  crucial step towards resolving its apparent lack of gauge invariance. We do this explicitly first for the black ring and then for the  black hole.

In  case of the black ring, even though we find that the  reduced action  is no longer invariant under large gauge transformations, it nevertheless  turns out  that the entropy function itself does remains gauge invariant.  Furthermore we show that this invariance is no coincidence, but stems from an underlying spectral flow symmetry of the theory, which leaves the entropy function invariant under spectral flow transformations. In order to achieve this, we have to first demonstrate how the relevant spectral flow relations emerge within the 5D  computation whilst solving the equations of motion in the presence of Chern-Simons terms.  Through this we shall also be able to identify the 4D/5D dictionary, using which the 4D-reduced computation of Goldstein and Jena  \cite{GJ}  can be  recovered - except for one subtle issue  on which our 5D computation differs from their 4D computation   for reasons that will become clear in  the calculations that follow.   

In this context it is worth pointing out to the work of \cite{MS} on $AdS$ black holes where it was also suggested that Chern-Simons terms would somehow facilitate charge shifts of the form $q_I \rightarrow q_I + c_I$. However these authors propose a modified Sen's formalism with the shifted charges directly implemented and $c_I$ are undetermined shift parameters. Here we take the viewpoint that it is not necessary to modify the formalism by imposing such a charge redefinition ad hoc, but rather a consistent 5D evaluation of Sen's functional should be possible from which these charge shifts emerge naturally. We will see that this is indeed the case due to 5D spectral flow. Moreover this way we are able to uniquely determine the shift parameters. 

After having treated the black ring, we next confirm  gauge invariance of the 5D black hole entropy function. Here again we see that a 5D calculation shows some interesting differences when compared  to the  4D calculations of \cite{GJ}. This will have something to do with the  $x^{\mu}$-dependence of the moduli $a^I$ ( which are  $\psi$-components of the 5D gauge fields $A^I$  ).  In the calculations of \cite{GJ}, the  $x^{\mu}$-dependence of  $a^I$ are retained throughout  dimensional reduction of  Chern-Simons terms to 4D and only then are they set as constants. Apparently this is what seems to create a seemingly incorrect  shift in electric charges when comparing their result for the  black hole entropy to that of \cite{L}.  Here we claim that the way out is not to assume such a coordinate dependence  (  which would even be incompatible with the isometries of the 5D near-horizon geometry  )  in a  5D calculation.  In addition to finding an agreement with the result of \cite{L}, our claim also leads to the correct 5D electric charges which are seen to perfectly tally with  recent results of \cite{HOT}, who  perform an explicit  near-horizon analysis of 5D supergravity. 

The outline of this note is as follows - In section 2 we compute the black ring entropy function without dimensional reduction. The 5D charges turn out to be Page charges, which exhibit spectral flow behaviour. The entropy function however  is shown to be spectral flow invariant. Section 3 concerns gauge invariance of the 5D  black hole entropy function.  For both black objects, we compare the 5D charges computed here via the 5D entropy formalism to those computed in the supergravity analysis of \cite{HOT}. In section 4 we provide an  interpretation for the  $e^0 \; \leftrightarrow \; p^0$  switch within the entropy formalism  as   corresponding  to  a black hole  $\leftrightarrow$  black ring interpolation in supergravity.  Finally in section 5 we conclude with some discussions.

\section{Spectral flow invariance of the black ring entropy function}  
Let us now perform a 5D computation of the black ring entropy function and derive the associated spectral flow relations from the equations of motion therein.

Consider the action of 5D minimal ungauged two-derivative supergravity theory coupled to  $N -1$ abelian vector multiplets. Writing  only the  bosonic fields, we have 
\begin{eqnarray}
 {\cal S}_5 = \frac{1}{16 \pi G_5} \int R \ast 1 - G_{IJ} d X^I \wedge \ast d X^J -   \frac{1}{2}  G_{IJ} F^I  \wedge \ast  F^J  -  C_{IJK} A^I \wedge F^J \wedge F^K  
\end{eqnarray}
where $X^I$ are massless scalars parameterizing  ``very special geometry''. These scalars  obey  the volume  constraint 
\begin{eqnarray}
C_{IJK} X^I X^J X^K = {\cal V}
\end{eqnarray}
The couplings $G_{IJ}$ are functions of the scalar moduli and are defined as
\begin{eqnarray}
G_{IJ} = - \frac{1}{2} \frac{\partial}{\partial X^I }  \frac{\partial}{\partial X^J } \; ln \, {\cal V} \; \bigg|_{ {\cal V} = 1  }
\end{eqnarray}
The indices $I$, $J$, $K$ run from $1$ to $N$ while  $C_{IJK}$  is a completely symmetric tensor and  $F^I = d A^I$ are $N$  $U (1)$ gauge fields.

The Chern-Simons term in the action above is clearly not invariant under large gauge transformations, such as 
\begin{eqnarray}
A^I  \longrightarrow  A^I + k^I
\label{22.4}
\end{eqnarray}
where $k^I$ are integral constants\footnote{Small gauge transformations pose no problems in this case since they do not affect the equations of motion. This is   because the extra gauge terms in the action can be expressed as an integral of a  total derivative which is then evaluated  as a surface term at infinity, where  the gauge  parameters asymptotically vanish.  However, with large gauge transformations this is not so.  }.  In this section, we revisit the black ring entropy function and show that instead of the 4D approach followed by  \cite{GJ}, one can also perform  an alternate well-defined 5D calculation.  Consequently,  the problematic Chern-Simons terms have now to be directly tackled; which we shall do by invoking spectral flow shifts.   

To begin with, the 5D geometry is expressed via a Kaluza-Klein ansatz for an  $AdS_2 \times S^{2} \times S^1$  topology  ( kindly refer to eq.(\ref{22.7}) below  ).  Eventually of course, when one extremises  the entropy function, the $S^1$ fibres  over the $AdS_2$  (  see \cite{GJ} ) precisely  recovering   the known  near-horizon $AdS_3 \times S^{2}$  metric  (  \cite{EEMR1}, \cite{EEMR2}, \cite{BK1}, \cite{B}  ) of a supersymmetric  black ring.
   
After  expressing the 5D gauge potential $A^I$ in terms of the aforementioned Kaluza-Klein  decomposition, we  implement gauge transformations of the type  in   eq.(\ref{22.4})  as  follows
\begin{eqnarray}
A^I =  A^I_{\mu} d x^{\mu} + \left( a^I  + k^I  \right) \left(    d \psi + A^0_{\mu} d x^{\mu}  \right)
\label{22.5}
\end{eqnarray}
where $\psi$ parametrises the $S^1$ circle with a periodicity of $4 \pi$;  the $A^0_{\mu}$ are off-diagonal entries in the 5D Kaluza-Klein metric ( which we shall write down shortly  ); the scalars $a^I$, which are  $\psi$-components of the 5D gauge potential $A^I$, are interpreted as axions in 4D; while  $A^I_{\mu}$ would just be the usual gauge potential in the four non-compact dimensions.  

The reduced action ( terminology not to be confused with dimensionally reduced action  ) is now defined by integrating the 5D lagrangian density over  $S^2 \times S^1$ - the spatial horizon of the black ring, spanned by $\theta$, $\phi$ and $\psi$
\begin{eqnarray}
{\cal F}_5^{br} =  \frac{1}{16 \pi G_5}  \int_{\Sigma} d \theta d \phi d \psi \sqrt{- g_5} {\cal L}_5
\end{eqnarray}
Our task  then is  to evaluate ${\cal F}_5^{br}$ in the background of the   Kaluza-Klein metric for an $AdS_2 \times S^{2} \times S^1$ near-horizon topology  
\begin{eqnarray}
ds^2 = \omega^{-1} \left[ v_1 \left( - r^2 dt^2 + \frac{d r^2}{r^2}  \right) +   v_2 \left( d \theta^2 + sin^2 \theta d \phi^2  \right) \right]  + \omega^{2} \left( d \psi + A^0_{\mu} d x^{\mu}  \right)^2
\label{22.7}
\end{eqnarray}
with  $A^I_{\mu}$ and $A^0_{\mu}$  specified by 
\begin{eqnarray}
A^I_{\mu} d x^{\mu} = e^I r d t + p^I cos \theta d \phi   \qquad  A^0_{\mu} d x^{\mu} =  e^0 r d t 
\end{eqnarray}
Here we take  $\omega$, $v_1$, $v_2$, $X^I$, ${a}^I$, $e^I$, $e^0$  to be  constants in the near-horizon geometry. These  will eventually be fixed as functions of the black ring  charges upon extremisation.   $\omega$ is the radius of the Kaluza-Klein circle; $v_1$, $v_2$ denote the $AdS_2$ and $S^2$ radii respectively; $p^I$ are magnetic charges and $e^I$ denote  the  corresponding  electric fields in 4D ( we shall soon write down the electric fields in 5D as well  ).   $e^0$ is  dual to the magnetic field associated to a $p^0$ charge ( or D6-brane charge  ).  However for rings, it is well known that the $p^0$ charge is   absent in the  immediate  vicinity  of  the   horizon.  In 4D,  $e^0$  too is treated as an electric field; however in 5D it will turn out to be associated to the angular momentum of the black ring along the $S^1$ direction.     

Putting all this together, and computing the 5D reduced action gives
\begin{eqnarray}
{\cal F}_5^{br} (  v_1, \,  v_2, \, \omega, \, X^I, \,  {a}^I, \, e^I, \, e^0, \, p^I   ) &=&  \left( \frac{2 \pi}{G_5} \right)  \left[  \, v_1 - v_2 +  \frac{ v_2 \; \omega^{3} \, ( e^0 )^2 }{ 4 \, v_1 }    \right.  \nonumber \\   -  \;   \frac{ v_1 }{  v_2 }   \; \omega \; \frac{G_{IJ}}{2} \, p^I p^J  &+&  \left. \frac{ v_2 }{  v_1 }  \; \omega \; \frac{G_{IJ}}{2}  \, ( e^I + \tilde{a}^I e^0 ) \; ( e^J + \tilde{a}^J e^0 ) \,  \right]   \nonumber \\  &+&     \left( \frac{ 24  \pi}{G_5} \right)  C_{IJK} \left[  \,  ( e^I + \tilde{a}^I e^0 ) \,  p^J \, \tilde{a}^K \,  \right]  
\label{2.7}
\end{eqnarray}
We get the three terms in the first line of eq.(\ref{2.7}) by computing   the five dimensional Ricci scalar; the second line comes from the 5D Yang-Mills term in the action; and the last line is obtained  from the Chern-Simons term.  It is important to note that this result here differs from that of \cite{GJ} on two counts\footnote{Our $G_{IJ}$ equals $2 f_{IJ}$ in the notation of \cite{GJ}. }  -  Firstly we have  shifts in the moduli  ${a}^I   \rightarrow   \tilde{a}^I \equiv ( {a}^I + k^I )$, which essentially encode large  gauge transformations in 5D and  consequently  leave  ${\cal F}_5^{br}$  with   a gauge ambiguity. In a 4D-reduced  calculation these shifts do not appear. The second point on which ${\cal F}_5^{br}$ differs from its dimensionally reduced version  ${\cal F}_4^{br}$ is a factor of $\frac{1}{2}$ in one of the two  Chern-Simons contributions to the reduced action.  In the next section we shall see that this difference of  factors  arises because of the way the moduli $a^I$ have to be treated in a 5D  calculation  as opposed to how they were dealt with in the 4D case.  This point will also turn out to be crucial in determining the correct 5D charges and in the end we shall justify our results by comparing with the analysis in \cite{HOT}.

Now, the 5D entropy function is defined as the Legendre transform of ${\cal F}_5^{br}$ with respect to electric charges $Q_I^{br}$, $Q_0^{br}$
\begin{eqnarray}
{\cal E}_5^{br} \; = \;  2 \pi \left[ \;  Q_0^{br} \,  e^0 \; + \; Q_I^{br} \cdot ( e^I + \tilde{a}^I e^0 ) \; - \; {\cal F}_5^{br} (  v_1, \,  v_2, \, \omega, \, X^I, \,  \tilde{a}^I, \, e^I, \, e^0  ) \;  \right]
\label{2.8}
\end{eqnarray}
where $Q_I^{br}$ and $Q_0^{br}$ are canonically conjugate to $( e^I + \tilde{a}^I e^0 )$ and $( e^0 )$ respectively 
\begin{eqnarray}
Q_I^{br}  =  \frac{G_5}{4 \pi} \;  \frac{\partial {\cal F}_5^{br} }{\partial ( e^I + \tilde{a}^I e^0 ) }   \qquad   \qquad    Q_0^{br}  =   \frac{G_5}{4 \pi}  \; \frac{\partial {\cal F}_5^{br} }{\partial e^0}
\label{2.9}
\end{eqnarray}
As we shall soon see, $Q_I^{br}$,  $Q_0^{br}$ are 5D Page charges and are physical  observables of the black ring. These charges  will differ from the 4D electric charges  $q_I$ respectively $q_0$ computed in  \cite{GJ}. 

Obtaining  the entropy of a black ring  then  entails   extremisation of  the   entropy  function ${\cal E}_5^{br}$  with respect to its moduli variables 
\begin{eqnarray}
\frac{\partial {\cal E}_5^{br} }{\partial a^I} =  \frac{\partial {\cal E}_5^{br} }{\partial v_1} =  \frac{\partial {\cal E}_5^{br} }{\partial v_2} =  \frac{\partial {\cal E}_5^{br} }{\partial \omega} =  \frac{\partial {\cal E}_5^{br} }{\partial X^I} = 0 
\label{2.10}
\end{eqnarray}
But before that let us see how the gauge ambiguity in the reduced action  ${\cal F}_5^{br}$, and consequently in the entropy function ${\cal E}_5^{br}$, can be resolved.   For that purpose we will need to know  exactly how the  Chern-Simons terms in ${\cal F}_5^{br}$ affect  physical  charges $Q_I^{br}$ and  $Q_0^{br}$. It turns out that they induce  spectral flow shifts in these  charges. And we want to know how these shifts can be manifestly derived  within the framework of the entropy function formalism itself.  Consequently we shall see how ${\cal E}_5^{br}$ remains invariant under these shifts.  

We begin by evaluating  eq.(\ref{2.9})  for $Q_I^{br}$ and  $Q_0^{br}$  by making use of   ${\cal F}_5^{br}$  from  eq.(\ref{2.7})
\begin{eqnarray}
Q_I^{br}  = \left(  \frac{v_2}{v_1} \right) \omega   \frac{G_{IJ}}{2}  \left( e^J + e^0  \tilde{a}^J \right)  +  6 C_{IJK} \tilde{a}^J p^K
\label{2.11}
\end{eqnarray}
and
\begin{eqnarray}
Q_0^{br}  =  \left(  \frac{v_2}{v_1} \right)  \left( \frac{1}{4} \omega^3 e^0 +  \omega  \frac{G_{IJ}}{2}  \tilde{a}^I  \left( e^J + e^0  \tilde{a}^J \right)  \right)  +   6  C_{IJK} \tilde{a}^I  \tilde{a}^J  p^K
\label{2.12}
\end{eqnarray}
That these are in fact the correct 5D charges for a black ring can be checked  by comparing these expressions  to the 5D Page charges recently computed in the supergravity analysis of \cite{HOT}, who showed that the near-horizon region of a black ring also encodes full information of its charges measured at asymptotic infinity.  The results of \cite{HOT} yield
\begin{eqnarray}
Q_I^{Page}  &=&  \frac{1}{64 \pi^2} \int_{\Sigma}  \; \ast F_I + 6 C_{IJK} A^J \wedge F^K 
\label{2.11a}   \\
Q_0^{Page}  &=&  - \frac{1}{64 \pi^2}  \int_{\Sigma}  \; \ast d \xi + \ast ( \xi \cdot A^I ) F_I + 6 C_{IJK} ( \xi \cdot A^I ) A^J \wedge F^K  
\label{2.12a}
\end{eqnarray}
where $\Sigma$  is a 3-cycle over the spatial horizon. For the black ring $\Sigma$  specialises to $S^2 \times S^1$.   $\xi$ denotes the axial Killing vector with respect to the $\psi$-direction, while $( \xi \cdot A^I )$ is an inner product between a vector field and a one-form. The Killing field $\xi$ generates isometries along the  $\psi$-direction; leading to a conserved charge,  which is simply the angular momentum. In fact, the right-hand side of eq.(\ref{2.12a}) is just the Noether charge of Wald.  Page  charges are in fact not gauge invariant ( due to an explicit $A^I$-dependence in these expressions  ), even though they are conserved and localised \cite{DM}.  Now  in order to  strike  a comparison between these  charges of  \cite{HOT}  and those computed here using  the 5D  entropy formalism, we will need to explicitly integrate the right-hand sides of  eqs.(\ref{2.11a}) and  (\ref{2.12a}). Since these are simply local integrations, it is sufficient to make use of only near-horizon data of the gauge fields and metric from  eqs.(\ref{22.5}) and  (\ref{22.7}).  Computing the non-vanishing components of the 5D field strength gives    $F^I_{rt} = e^I + \tilde{a}^I e^0$  and $F^I_{\theta \phi} = - p^I sin \theta$. In the near-horizon terminology, the axial vector $\xi^i$ is found to be $A_0^i$, with non-vanishing components $A_0^t = \frac{\omega^3 e^0}{v_1 r}$ and $A_0^{\psi} = -1$. Using this  we can determine $F_0$, which is just $ d \xi$; and by  $d \xi$ we mean $\partial_i \xi_j \; dx^i \wedge dx^j$. Note also that in the $\xi \cdot A^I$ term,  it makes sense to only  consider the projection of the Killing field on the physical (  on-shell sector ) gauge fields.  Putting together all these quantities  and inserting them  into  eqs.(\ref{2.11a}) and  (\ref{2.12a}) exactly reproduces eqs.(\ref{2.11}) and  (\ref{2.12}). Hence  we see that  $Q_I^{br}$ and $Q_0^{br}$ obtained from the entropy function indeed represent the correct five-dimensional supergravity Page charges  $Q_I^{Page}$ and $Q_0^{Page}$  respectively.  

Now in the entropy function formalism the 5D field $A^I$ in eq.(\ref{22.5}) depends on three different  moduli  $e^I$, $e^0$ and $a^I$.  Extremising  ${\cal E}_5^{br}$  with  respect to these moduli and plugging the extremum values of these moduli back into eq.(\ref{22.5}) basically determines the near-horizon gauge fields of the black ring. $A^I$ can then be expressed purely in terms of electric and magnetic charges.  For our purposes, these three extremisation conditions will fully determine the physical charges that source these gauge fields $A^I$.   Hence  eqs.(\ref{2.11}) and  (\ref{2.12})  require further input from
\begin{eqnarray}
\frac{\partial {\cal E}_5^{br} }{\partial a^I} = 0   \qquad  \Longrightarrow  \qquad   F^I_{rt} = 0
\label{2.13}
\end{eqnarray}
and this exactly corresponds to $\int_{\Sigma}  \; \ast F_I = 0$  computed in \cite{HOT} by explicit near-horizon integration. Physically, eq.(\ref{2.13})  signifies a vanishing electric flux in the near-horizon geometry, which is simply    what one would expect in the absence of a compact 3-cycle when the topology is that of  $AdS_3 \times S^2$.

We are now ready to identify the black ring spectral flow shifts that  emerge from within  the structure of the entropy function formalism itself.   Separating  the  $k^I$  dependence in $Q_I^{br}$ and   $Q_0^{br}$  yields 
\begin{eqnarray}
Q_I^{br} =  q_I + 6 C_{IJK} {k}^J p^K  
\label{2.14}
\end{eqnarray}
and
\begin{eqnarray}
Q_0^{br}  =   q_0  +  k^I   q_I  +  6  C_{IJK} {k}^I   {k}^J  p^K  
\label{2.15}
\end{eqnarray}
where $q_I$ and $q_0$  are read-off  from eqs.(\ref{2.11})  respectively  (\ref{2.12}) after replacing $\tilde{a}^I$  by ${a}^I$; and they can indeed be identified as  the four dimensional (  gauge invariant as well ) electric charges that appeared in the calculation of \cite{GJ}. In 5D however,   $Q_I^{br}$ and  $Q_0^{br}$ are the correct physical observables \cite{BKW}, \cite{HOT}, \cite{XDA}.

Let us now  determine what the conserved quantities, under spectral flow shifts  of   $Q_I^{br}$ and  $Q_0^{br}$  look like. It is easy to see that  $\widehat{Q}_0$ defined by
\begin{eqnarray}
\widehat{Q}_0  \; \equiv \;  Q_0^{br} - \frac{1}{2} C^{IJ} Q_I^{br} \; Q_J^{br}
\label{2.17}
\end{eqnarray}
is left invariant  under  spectral flow transformations described in  eqs.(\ref{2.14})  and  (\ref{2.15})  in the following sense 
\begin{eqnarray}
\widehat{Q}_0  \left( Q_0^{br} , \;  Q_I^{br}  \right)  \; =  \;  \widehat{Q}_0   \left( q_0 , \; q_I   \right) 
\label{2.171}
\end{eqnarray}
where   $C^{IJ} \equiv \left[ C_{IJ} \right]^{- 1}$  and $C_{IJ} \equiv 6 C_{IJK} p^K$.    This invariant $\widehat{Q}_0$   is in fact  the $E_{7(7)}$  quartic invariant of \cite{KK} and  will play a role in  maintaining  invariance of the  5D black ring entropy function.

Putting together all the above  ingredients into eq.(\ref{2.8})  gives us  the entropy function in terms of 5D variables 
\begin{eqnarray}
{\cal E}_5^{br} =   \frac{4 \pi^2}{G_5}  \left\{ v_2 - v_1  +  \frac{v_1}{v_2}  \left[   \omega  \frac{G_{IJ}}{2}  p^I p^J   +   4     \omega^{-3}  \left(  \widehat{Q}_0    \right)^2    \right]  \right\}
\label{2.19}
\end{eqnarray}
The first term  in the square brackets  in  ${\cal E}_5^{br}$ comes from the  magnetic flux, while  the second term is related to the effective  momentum  of D0-particles\footnote{ These are precisely the left-movers of the dual $(0,4)$ SCFT \cite{MSW}. }.  This  brings us to the  main result of this section   that ${\cal E}_5^{br}$  is  indeed  invariant under spectral flow transformations, once the moduli  of the  gauge field $A^I$ have been  determined.  Here we have obtained  ${\cal E}_5^{br}$  in eq.(\ref{2.19})  from a 5D calculation, and this agrees with the dimensionally reduced ${\cal E}_4^{br}$ of \cite{GJ} due to  spectral flow invariance\footnote{ The 4D/5D lift for black rings is  in fact  a special case of spectral flow transformations when the value of $k^I$ is set to $p^I$ \cite{XDA}.  }.  Remarkably the final 4D and 5D entropies reconcile despite the fact that  ${\cal F}_5^{br}$ differs from  ${\cal F}_4^{br}$ as does the definition of charges.  This illustrates  the point that for a 5D action which includes Chern-Simons terms, there is another way besides  dimensional reduction to  4D;  a direct 5D calculation will also  lead to the correct result once the correct 5D variables have been  implemented into the calculation.   Note that  ${\cal E}_5^{br}$ is   not  yet an  entropy and here  what we see is  that even when  ${\cal E}_5^{br}$ is not at its stationary point, it is  still  gauge invariant. Hence  we  get     
\begin{eqnarray}
{\cal E}_5^{br}  \left( Q_0^{br} , Q_I^{br} , p^I , v_1 , v_2 , \omega , X^I   \right)  =  {\cal E}_5^{br}  \left( q_0 , q_I , p^I , v_1 , v_2 , \omega , X^I   \right)
\label{2.20a}
\end{eqnarray}
upon inserting eqs.(\ref{2.14})  and  (\ref{2.15}) into eq.(\ref{2.19}). The left-hand side is what one gets from an explicit 5D calculation, whereas the right-hand side is what results from a dimensionally reduced computation.   

A 5D calculation is necessary to illustrate the inherent  spectral flow associated to a black ring geometry.  The physical interpretation of  spectral flow for  black rings has been discussed in \cite{XDA}.  The 4D/5D transformations themselves  are in fact a special case of  spectral flow transformations.  And that is actually the reason why  application of the  entropy function   formalism to black rings should work well either in 4D or 5D ( even though we think that an explicit 5D computation expresses charge/geometric data more naturally  ).

For the sake of completeness, let us also extremise with respect to the remaining  moduli, as  in eq.(\ref{2.10}); and show that the resulting black ring entropy obtained from our  5D  calculation  indeed  gives the right answer. Solving for   $v_1$, $v_2$, $\omega$  gives 
\begin{eqnarray}
v_1 \; = \; v_2 \; = \;  \omega \frac{G_{IJ}}{2}  p^I p^J  +   4 \omega^{- 3} ( \widehat{Q}_0 )^2   
\label{2.24}
\end{eqnarray}
and
\begin{eqnarray}
\omega^4  \;  =  \;  \frac{12 \; ( \widehat{Q}_0 )^2 }{ \frac{G_{IJ}}{2} \,  p^I p^J  }
\label{2.25}
\end{eqnarray}
and upon using these  values of  $v_1$, $v_2$, $\omega$  back into  ${\cal E}_5^{br}$  yields 
\begin{eqnarray}
{\cal E}_5^{br} = \frac{8 \pi^2}{G_5}  \sqrt{ \left(  \frac{2 \; G_{IJ}}{3}  p^I p^J   \right)^{\frac{3}{2}}  \widehat{Q}_0  }
\label{2.26}
\end{eqnarray}
Of course the couplings $G_{IJ}$, which are  functions of the yet-to-be-extremised  scalar moduli $X^I$, will depend on  geometric data  of the specific compactification space. For our purposes we leave it with the general expression in eq.(\ref{2.26}).

\section{  Gauge invariance of the 5D black hole entropy function  }
We now repeat our calculation for the 5D  black hole. The near-horizon metric ansatz is again taken to be  $AdS_2 \times S^2 \times S^1$. However this time round it turns out that  the $S^1$ fibres over the $S^2$, eventually leading to an $AdS_2 \times S^3$ geometry near the horizon \cite{GJ}.  It has been proven in \cite{KLR}, \cite{AY} that even in the rotating case, the near-horizon isometry of an extremal black hole contains an $SO(2,1)$ symmetry.  Moreover, that the entropy function formalism can also be applied to such rotating black holes having $AdS_2$ isometry was shown in \cite{AGJST}.  Such a black hole in 5D carries a Kaluza-Klein monopole charge, which comes from  uplifting  a D6-brane in  Type II A  theory to M-theory and the black hole sits at the origin of the KK monopole. Even though this geometric configuration is different from that of a black ring,  it is  still reasonable to  implement the Kaluza-Klein  metric ansatz of  eq.(\ref{22.7})  provided  the off-diagonal components $A^0_{\mu}$  are suitably modified for the black hole case. We consider the same type of black hole as in  \cite{GJ}, so that the results of our analysis can be compared to theirs. Hence $A^0_{\mu}$ is taken as
\begin{eqnarray}
A^0_{\mu} d x^{\mu} =  p^0 \, cos \, \theta \, d \phi 
\label{3.1}
\end{eqnarray}
where $p^0$ denotes the Kaluza-Klein monopole charge. Note also that  the quantity  $e^0$  is not defined  for these black holes, which  corresponds to an absence of Kaluza-Klein momentum $J^{KK}_0$.  Here  $J^{KK}_0 = 0$ is only to be thought of as vanishing of the intrinsic angular momentum ( resulting from the absence of D0-charge in the brane bound state ). In \cite{GJ} it was claimed that this black hole is static. However there is a slight subtlety to that\footnote{We are grateful to Dumitru Astefanesei for a discussion on this point.}.  The effective angular momentum is in fact non-vanishing. As a quick check one can easily compute the integral in  eq.(\ref{2.12a}) and we see that the second term in the integrand carries a non-vanishing contribution. Nevertheless it will turn out that this effective contribution does not enter the entropy formula ( and this last point was presumably the reason that this black hole was viewed as a static system in \cite{GJ} ).  On the other hand a black hole of the BMPV type \cite{BMPV}, is a true rotating black hole with an angular momentum that enters the entropy formula. Such a black hole would be obtained had we started with a bound state of spinning M2's in Taub-NUT ( or a D0-D2-D6 bound state in Type II A ). Instead what we have here is a black hole more of the type discussed in \cite{GMT}.  It can be conceived as a  bound state of non-rotating M2's sitting at the tip of a Taub-NUT-flux geometry ( D2-D4-D6 in II A ), where the intrinsic angular velocity of the horizon vanishes, leaving only the flux induced component of the angular momentum which affects the geometry but not the entropy formula - in some sense like a static black hole in a flux background. 

Within this set-up we now compute  ${\cal F}_5^{bh}$ to get 
\begin{eqnarray}
{\cal F}_5^{bh} (  v_1, \,  v_2, \, \omega, \, X^I, \,  {a}^I, \, e^I, \, p^I, \,  p^0   ) &=&  \left( \frac{2 \pi}{G_5} \right)  \left[  \, v_1 - v_2 -  \frac{ v_1 \; \omega^{3} \, ( p^0 )^2 }{ 4 \, v_2 }    \right.  \nonumber \\  +  \;  \frac{ v_2 }{  v_1 }  \; \omega \; \frac{G_{IJ}}{2}  \,  e^I  \;  e^J    &-&  \left.    \frac{ v_1 }{  v_2 }   \; \omega \; \frac{G_{IJ}}{2} \, ( p^I + \tilde{a}^I p^0 ) \; ( p^J + \tilde{a}^J p^0 )   \,  \right]   \nonumber \\  &+&     \left( \frac{ 24  \pi}{G_5} \right)  C_{IJK} \left[  \,  ( p^I + \tilde{a}^I p^0 ) \,  e^J \, \tilde{a}^K \,  \right]  
\label{3.2}
\end{eqnarray}
which differs from  eq.(\ref{2.7}) with the replacement $p^I \longrightarrow  p^I +  \tilde{a}^I p^0$ and a $( p^0 )^2$  term in  the 5D Ricci scalar that replaces the $( e^0 )^2$ term in the black ring computation. Just as in the black ring analysis before, we once again find that  ${\cal F}_5^{bh}$ computed here is not exactly going to be the same as  ${\cal F}_4^{bh}$ in \cite{GJ}.  Firstly, in a 5D approach the gauge parameters $k^I$ show up and secondly, the relative factors in front of the Chern-Simons contributions will differ from those in the  4D computation of \cite{GJ}. We shall consequently see that this will definitely  affect the definition of electric charges in 5D and thereby fix a small mismatch, with respect to the definition of 5D charges, in the result for the entropy obtained by \cite{GJ} with that of \cite{L}. 

Having  eq.(\ref{3.2}) in hand, we are now in a position to write  the 5D  black hole charges  from the analog of the definition in eq.(\ref{2.9})
\begin{eqnarray}
Q_I^{bh}  = \left(  \frac{v_2}{v_1} \right) \omega   \frac{G_{IJ}}{2}  e^J   +  6 C_{IJK}  \left( p^J +  \tilde{a}^J p^0  \right)  \tilde{a}^K
\label{3.3}
\end{eqnarray}
Moreover using 
\begin{eqnarray}
\frac{\partial {\cal E}_5^{bh} }{\partial a^I} = 0   \qquad  \Longrightarrow  \qquad   p^I + \tilde{a}^I  p^0 = 0
\label{3.4}
\end{eqnarray}
we can  write  eq.(\ref{3.3})  as 
\begin{eqnarray}
Q_I^{bh} =  \int_{\Sigma}  \; \ast F_I
\label{3.5}
\end{eqnarray}
since $F^I_{rt} = e^I$ and $F^I_{\theta \phi} = - ( p^I + \tilde{a}^I  p^0 ) sin \theta$.   Here $\Sigma$ is now an $S^3$, the spatial horizon of the  black  hole.   Eq.(\ref{3.4})  is just the condition for vanishing of the effective magnetic flux
\begin{eqnarray}
\int_{S^2}  \;  F_I \; = \; 0
\label{3.6}
\end{eqnarray}
in other words suggesting the absence of a compact 2-cycle in this black hole geometry.  Moreover for given magnetic charges $p^I$ and $p^0$,  the constraint $p^I + \tilde{a}^I  p^0 = 0$ imposes a restriction on the value of $k^I$. Therefore for this black hole, we cannot set-up arbitrary spectral flow shifts for the charges. 

In the terminology of \cite{DM}, eq.(\ref{3.5}) implies that  $Q_I^{bh}$ is not a Page but a Maxwell charge\footnote{  Additionally, in this case the Maxwell charge is localised within $\Sigma$  and does not require integration over all space because the source term $F^J \wedge F^K$ in the 5D  supergravity equation of motion :  $d \ast F_I = - 6 C_{IJK} F^J \wedge F^K$,  vanishes following  eq.(\ref{3.4}).    }, which is gauge invariant and  does not show  spectral flow behaviour.  $Q_I^{bh}$ therefore represents the same physical observable in 5D as well as in 4D alike.  

Under these considerations, the entropy function for this black hole takes the form
\begin{eqnarray}
{\cal E}_5^{bh} =   \frac{4 \pi^2}{G_5}  \left\{ v_2 - v_1  +  \frac{v_1}{v_2}  \left[  \frac{1}{4}  \omega^3  ( p^0 )^2   +     \omega^{-1} 2 G^{IJ}  Q_I^{bh}  Q_J^{bh}     \right]  \right\}
\label{3.7}
\end{eqnarray}
where $G^{IJ}$ is defined as the inverse of $G_{IJ}$.  Once again we have obtained a gauge invariant  entropy function from an explicit 5D calculation in terms of physical 5D variables.  Now it is straightforward to extremise ${\cal E}_5^{bh}$ with respect to $v_1$, $v_2$ and $\omega$ to get
\begin{eqnarray}
v_1 \; = \; v_2 \; = \;   \frac{1}{4}  \omega^3  ( p^0 )^2   +     \omega^{-1} 2 G^{IJ}  Q_I^{bh}  Q_J^{bh}    
\label{3.8}
\end{eqnarray}
and
\begin{eqnarray}
\omega^4  \;  =  \;  \frac{ 8 G^{IJ}  Q_I^{bh}  Q_J^{bh}  }{ 3 ( p^0 )^2   }
\label{3.9}
\end{eqnarray}
Then eliminating  $v_1$, $v_2$ and $\omega$  by way of substituting their values at the stationary point back into   ${\cal E}_5^{bh}$  leaves us with 
\begin{eqnarray}
{\cal E}_5^{bh} = \frac{4 \pi^2}{G_5}  \sqrt{ p^0 \; \left(  \frac{ 8 \; G^{IJ} }{3}  Q_I^{bh}  Q_J^{bh}   \right)^{\frac{3}{2}}   }
\label{3.10}
\end{eqnarray}
which finally gives us the entropy of this black hole. The couplings $G^{IJ}$ can be determined  depending  on the specific choice of  compactification.  Here $Q_I^{bh}$  is the observable electric  charge in 5D  and since we have shown above that this charge does not exhibit  any spectral flow behaviour, it exactly equals the number of M2-branes wrapping Calabi-Yau 2-cycles. Upon shrinking the M-theory circle and reducing  to Type II A, the M2-branes directly descend to D2-branes. Then  $Q_I^{bh}$  is also the physical charge for a 4D black hole. 

Now comparing  eq.(\ref{3.10}) above to the entropy obtained by \cite{L} ( whose  computation is performed via a 5D attractor mechanism ), gives an exact agreement.    This fixes the mismatch in the  result of \cite{GJ} where the charges  $Q_I^{bh}$ in the entropy formula  were shifted by  $3C_{IJK}p^Jp^K/p^0$.  In our case, using   eqs.(\ref{3.3}) and (\ref{3.4})  we see that the charges entering the entropy are $Q_I^{bh}  = \left(  \frac{v_2}{v_1} \right) \omega   \frac{G_{IJ}}{2}  e^J$  without any $p^I$ dependence.  The extra  $3C_{IJK}p^Jp^K/p^0$  terms in  \cite{GJ} stem from a 4D reduction of  Chern-Simons terms. This is what gives the authors of \cite{GJ} a relative factor of $\frac{1}{2}$ within the  Chern-Simons parts of ${\cal F}_4^{bh}$. In our 5D calculation this factor does not appear. And that has to do with how we treated the moduli $a^I$ in our calculation, as opposed to how the same  have been handled in  \cite{GJ}. To start with, let us verify the validity of the electric charges computed here. We want to check whether our $Q_I^{bh}$ compares to the charge integral obtained in the supergravity analysis of \cite{HOT}, which serves as an independent check. For that purpose consider  eq.(\ref{2.11a})  with $\Sigma$ taken to be an  $S^3$.  Since we know the near-horizon components of $A^I$ and $F^I$, we  insert these into  eq.(\ref{2.11a})  and evaluate the integral.  Because  $F^I_{\theta \phi} = 0$, the  $\int_{\Sigma}  A^J \wedge F^K$ part of the integral  vanishes  and the  $\int_{\Sigma}  \ast F_I$  term precisely  reproduces $\left(  \frac{v_2}{v_1} \right) \omega   \frac{G_{IJ}}{2}  e^J$.  That verifies our expression for the black hole electric charge.  One may now ask why the  charges of \cite{GJ} picked up those incorrect shifts ? Which may  be rephrased by asking what went wrong with their  Chern-Simons  contributions to ${\cal F}_4^{bh}$ ?   The problem seems to arise because they assume an $x^{\mu}$-dependence for the moduli $a^I$, while performing a dimensional reduction  of Chern-Simons terms. These  $a^I$ are set to constants only when one arrives at the four dimensional set-up. However this introduces  terms in ${\cal L}_{CS}^{4D}$, which wrongly shift the electric charges thereby causing a mismatch with the entropy of \cite{L}.  On the other hand,  in our 5D calculation, in the absence of any dimensional reduction there is no natural way to assume an $x^{\mu}$-dependence for  $a^I$ ( whilst already in the 5D near-horizon geometry ) and then suddenly set them to constants at some other stage of the calculation. The 5D components of the field strength $F_{rt}$, $F_{\theta  \phi}$ are constants in the near-horizon geometry and giving the fields  an $x^{\mu}$-dependence through  $a^I$ would also seem to come in conflict with the isometries of the near-horizon geometry.  Therefore in our calculations we have set all 5D near-horizon moduli as constants ( whose values are determined upon extremisation  ) throughout the calculation and this seems to give the correct answer.

\section{\texorpdfstring{Switching roles of $e^0 \leftrightarrow p^0$ as a form of black ring - black hole interpolation}{Switching roles of e(0) - p(0) as a form of black ring - black hole interpolation}}
Earlier in section 2 we saw how the near-horizon solution of a black ring can be  expressed   via  various moduli parameters.  Among these  $e^I$ and $e^0$ are conjugate to the electric charges and angular momentum respectively, while the magnetic flux $p^I$  is a fixed quantity. On the other hand, the 5D  black hole of section 3 only carried electric variables $e^I$ and fixed magnetic variables $p^I$, $p^0$.  From the perspective  of  the entropy function formalism, obtaining the metric of a black hole from that of a black ring can simply be achieved by switching off the $e^0$ contribution to  the metric and turning on a $p^0$ one  instead  ( and then extremising with respect to these new moduli ).  This assignment was first proposed in  \cite{GJ}, where it appears  as an ad hoc choice that  reproduces the leading order entropies of the two black objects. In this section we want to provide  a physical justification  for this assignment of parameters.  We will soon see  that switching the terms $e^0 rdt$ $\leftrightarrow$ $p^0 cos \theta d \phi$ among each other  in the near-horizon Kaluza-Klein metric will in fact be equivalent to changing the modulus $l$  (  here $l$ is the three dimensional distance of the black ring from the origin of the Taub-NUT base space  )  from a specified finite quantity to a  vanishing limit in the complete 5D supergravity solution.  Gravitationally this  means we are  shrinking the 5D   black ring to the origin of the base space to get a  5D black hole. In this sense, we argue that the  $e^0 \leftrightarrow p^0$ switch  is actually a  black hole - black ring  interpolation rather than some sort of black hole - black ring duality, that  was suggestively  speculated in \cite{GJ}.  Let us now examine this in more detail.

In section 2 we demonstrated that $p^I$, $Q_I$ and $Q_0$ computed from a 5D  entropy function analysis, are the correct  physical observables of a black ring. Moreover a glance at the  microscopic description of a black ring as a bound state of branes will in fact reveal  that the observable charges are  not exactly   the brane charges \cite{BW1}, \cite{BK3}.   Microscopically a black ring can be described by a Calabi-Yau compactification of M-theory on a circle  \cite{CGMS}    with M2-M5 branes wrapping 2- respectively 4-cycles on the  Calabi-Yau.  The remaining one leg of the M5-brane wraps the M-theory circle thus giving a black string along this $S^1$  ( as in the description  of \cite{MSW}  ).  This string is stabilised   by angular momentum modes running along the circle.  The relation between  brane charges and  observable charges in fact takes the form    \cite{BKW}    
\begin{eqnarray}
q_I^{M2} &=& Q_I - 6 C_{IJK}  \; p^J_{M5} \; p^K_{M5}   \nonumber  \\
p^I_{M5} &=& p^I  \nonumber  \\
J_0^{KK} &=& Q_0 - p^I_{M5}  \; q_I^{M2} -   6 C_{IJK}  \; p^I_{M5}  \; p^J_{M5}  \;  p^K_{M5}  
\label{4.1}
\end{eqnarray}
These  shifts from the actual brane charges have been shown in \cite{XDA} to be manifestations of spectral flow when  $k^I = p^I$. In this way the above relations also serve as a 4D/5D map between the two-center system of a  D0-D2-D4 black hole in 4D, placed in the vicinity of a D6-charge; and a black ring in 5D.  Hence  when the M-theory circle shrinks to zero size then the charge shifts due to spectral flow disappear and the brane charges $J_0^{KK}$, $q_I^{M2}$, $p^I_{M5}$  ( which  now become  D0, D2, D4 charges respectively in the Type II A description  ) coincide with the observable charges. Having stated the relations  between  physical and brane charges of the black ring, we can now incorporate these into  supergravity solutions.

In order to study a supergravity construction that interpolates between 5D black holes and black rings in its different limits, we start by considering  the most general 5D ${\cal N} = 1$ ungauged supergravity solution  \cite{GGHPR},  \cite{GG3}    which is given  by the  following 5D metric and  gauge fields
\begin{eqnarray} 
ds^2_{5} &=& -  \; f^2  \; (  \; dt \; + \;  \Omega  \; )^2 \;  + \;  f^{-1} \; ds^2 ( M_4 )  \nonumber \\
F^I &=&  d \left[ \; f  \; X^I \; ( \; dt  \; +  \; \Omega \; )  \;  \right]  \; -  \;  \frac{2}{3}  \; f \; X^I  \;   (  \; d \Omega   \; +   \; \star \, d \Omega  \; ) 
\label{4.2}
\end{eqnarray}
where \;  $X^I$ \; are scalar fields in abelian vector  multiplets. They satisfy the constraint equation  $C_{IJK} X^I X^J X^K =1$ and $X_I$ are defined by the condition  $X^I X_I = 1$.   $ds^2 ( M_4 )$  above refers to the Gibbons-Hawking metric of a 4D hyper-Kahler base space, which in our case is simply taken to be  $ds^2 ( TN )$, the Taub-NUT metric ( or  $ds^2 ( {\mathbb R}^4 )$  when considering  a black ring in flat space  )  having  KK-monopole charge.    Let $r$, $\theta$, $\phi$,  $\psi$ denote coordinates on the 4D base space with $( r, \theta, \phi )$ locally parameterising  an  ${\mathbb R}^3$ and $\psi$ running  along a compact $S^1$ with periodicity $4 \pi$.  The Hodge dual $\star$ is taken with respect to the 4D base space.   The function $f$ and the one-form $\Omega$ can then be determined  in terms of four   harmonic functions  $H_{TN}(x)$, $K^I(x)$, $L_I(x)$ and $M(x)$ ( with $x \in  {\mathbb R}^3$ ) in the following sense
\begin{eqnarray} 
f^{-1}  \; X_I  &=&  \frac{1}{4}  \;  {H_{TN}}^{-1} \; C_{IJK} K^J K^K   \; +  \;  L_I   \nonumber \\
\Omega &=& (  \; -  \frac{1}{8} \;  {H_{TN}}^{-2}  \;  C_{IJK} K^I K^J K^K   \;  -  \;   {H_{TN}}^{-1}  \;  L_I  \; K^I  \;  +  \;  M   \; )   \nonumber \\   &\times&   (  \; d \psi  \;  +  \;  cos \; \theta  \; d \phi  \;  )  \;  +  \;  \widehat{\Omega}
\label{4.3}
\end{eqnarray}
where  $\widehat{\Omega}$ is defined by
\begin{eqnarray} 
\nabla \times \widehat{\Omega}   \; =  \;  H_{TN}  \; \nabla M   \; -  \;  M  \;  \nabla H_{TN}  \; +  \;   K^I  \; \nabla  L_I  \;  -  \;  L_I  \; \nabla K^I
\label{4.4}
\end{eqnarray}
Operating  the gradient on both sides of  this equation yields  integrability conditions
\begin{eqnarray} 
  H_{TN}  \; \nabla^2 M   \; -  \;  M  \;  \nabla^2 H_{TN}  \; +  \;   K^I  \; \nabla^2  L_I   \;  -  \;   L_I  \; \nabla^2 K^I    \; =  \; 0
\label{4.5}
\end{eqnarray}
which are  evaluated at each pole ( charge center ) in ${\mathbb R}^3$.  

Within the above framework, a supergravity solution for any black object is now  reduced to the task of specifying  four harmonic functions. Let us first write these down for a black ring and then we shall see how to interpolate them to a black hole solution. For a black  ring we have  the following
\begin{eqnarray} 
H_{TN} (x) \; \; =  \; \; \frac{4}{R^2_{TN}} + \frac{p^0_{KK}}{|x|}  \qquad  \qquad   
L_I (x) &=&  v_I + \frac{ q_I^{M2} }{|x - l|}   \nonumber \\
K^I (x)  \; \; =  \; \; \frac{ p^I_{M5} }{|x - l|} \qquad \qquad \quad  \; \; 
M (x) &=&   v_0  + \frac{  J_{0}^{KK} }{|x - l|} 
\label{4.6}
\end{eqnarray}  
Here $p^0_{KK}$ is the charge of the Kaluza-Klein monopole in M-theory, which  reduce to  $p^0_{KK}$ D6-branes in Type II A. The case $p^0_{KK} = 1$ corresponds to a  Taub-NUT, otherwise the  4D hyper-Kahler base space is an orbifold of Taub-NUT, such that its geometry in the neighbourhood of the origin is of the type ${\mathbb C}^2/{\mathbb Z}_{p^0_{KK}}$.  Let us clarify the  remaining notation as well :  $R_{TN}$ denotes the asymptotic radius of the original Taub-NUT; $x \in {\mathbb R}^3$  and  $l$ is a modulus in ${\mathbb R}^3$ which denotes the distance between the plane containing the $S^1$ of the ring and the origin of base space.  $v_I$ is a constant determined at infinity and $v_0$ will soon get fixed via the integrability conditions. These harmonic functions have been  specified  via  brane charges in the system. The bound states of branes wrapping Calabi-Yau cycles  form BPS point particles in ${\mathbb R}^3$  and the poles in the above harmonic  functions are attained precisely at the location of these BPS particles. The  M2-M5-$J^{KK}$ particle sits at $x = l$, while the KK monopole  is located at $x = 0$.  From a 4D point of view  this is a 2-center black hole system, but in 5D it's just a black ring in a Taub-NUT orbifold  \cite{GSY2}. 

Now let us evaluate eq.(\ref{4.5}) for the above harmonics at each of the two poles. This yields the following two  integrability conditions
\begin{eqnarray}
v_0 &=&  - \frac{J_0^{KK}}{| l |}   \label{4.7}   \\
J_0^{KK} &=& v_I \, p^I_{M5} \left( \frac{p^0_{KK}}{| l |}  +   \frac{4}{R^2_{TN}}  \right)^{-1}
\label{4.8}
\end{eqnarray}
Physically this implies  that  $J_0^{KK}$; which contributes part of the angular momentum along the $\psi$-direction of the ring;  cannot be arbitrarily chosen, but is fixed for a given configuration.  The above  conditions can then be  inserted  back into eq.(\ref{4.6}) and thereafter implementing  the charge transformations in eq.(\ref{4.1}) ( which were obtained as spectral flow shifts from the supergravity action ), essentially lays down the complete black ring solution.   This  compares to the  standard solutions of \cite{EEMR1},  \cite{EEMR2},  \cite{BK1}, \cite{B}, \cite{BKW}  when expressed in more convenient coordinates - but we will not require that here. 

Now let us study the behaviour of this  black ring  in the limit   $l \rightarrow 0$.  From \cite{EEMR3} we already know that we should  recover a 5D black hole in this limit. However the purpose of our presentation is to make a clear distinction  between   branes that constitute a black ring bound state  from those that constitute a black  hole bound state when the modulus $l$ is driven to zero.
 Then we want to relate these brane charges to the spectral flow of those  respective black objects in order to determine the physical charges. 

Let us begin with eqs.(\ref{4.7}) and  (\ref{4.8}). When  $l \rightarrow 0$, they  reduce to 
\begin{eqnarray}
J_0^{KK} &=&  0   \label{4.9}   \\
v_0 &=&  -  \frac{v_I \; p^I_{M5} }{ p^0_{KK} }
\label{4.10}
\end{eqnarray}
and the harmonics in eq.(\ref{4.6}) become
\begin{eqnarray} 
H_{TN} (x) \; \; =  \; \; \frac{4}{R^2_{TN}}  +  \frac{p^0_{KK}}{|x|}  \qquad  \qquad   
L_I (x) &=&  v_I + \frac{ q_I^{M2} }{|x|}   \nonumber \\
K^I (x)  \; \; =  \; \; \frac{ p^I_{M5} }{|x|}  \qquad \qquad  \qquad \; \; \;
M (x) &=&   -   v_I \; p^I_{M5}
\label{4.11}
\end{eqnarray}  
after having used  eqs.(\ref{4.9}) and  (\ref{4.10}) therein. What we have now is  a BPS configuration in which there is not only a KK monopole  at the origin of the Taub-NUT orbifold, but also the M5-M2 charge is now bound to this monopole.  Moreover these bound states of branes  have  vanishing $J_0^{KK}$ charge.   This is a 5D black hole   ( or a D2-D4-D6 black hole from the point of view of a  4D reduction  ).   Furthermore from the analysis in section 3 we saw that in the case of the 5D black hole,  there are no  spectral flow shifts.  Therefore for this configuration, the brane charges $p^0_{KK}$, $p^I_{M5}$  and   $q_I^{M2}$  respectively  correspond to  the following  physical charges
\begin{eqnarray}
p^0_{KK} &=&  p^0  \nonumber  \\
p^I_{M5} &=&  p^I  \nonumber  \\
q_I^{M2} &=&  Q_I 
\label{4.12}
\end{eqnarray}
Now recalling the entropy function formalism, these  charges are precisely  associated to the following  near-horizon  variables : $p^0$, $p^I$, $e^I$. To sum up the contents of this section, we find that the  physical interpretation of switching  $e^0$ with $p^0$ in the entropy formalism's  near-horizon ansatz  corresponds to interpolating  between  limits of the modulus  $l$ on a  Taub-NUT orbifold, which in supergravity yields an interpolation between black hole/black ring geometries.  Moreover building this association to  supergravity also serves the purpose of providing a  justification for the specific choice of  moduli  in the Kaluza-Klein metric ansatz of  \cite{GJ},  for each of the two geometries.

\section{Discussion and conclusions}  
The inclusion of Chern-Simons terms in the entropy function formalism has rather been a bit of a puzzle  due to its apparent lack of gauge invariance under large gauge transformations. This being because Sen's original derivation \cite{S1} was based on the premise of gauge and reparametrisation invariant lagrangian densities.  The  dimensional  reduction approach was proposed \cite{SS1} in order to  rectify  this.  In view  of the proposed 4D/5D connection \cite{GSY1}, \cite{GSY2},  that such a recipe works might not come as a total surprise though. However even in those developments several contentious subtleties stood out as regards the correct physical notion of charge in 4D and 5D \cite{BK3},  \cite{BKW}, \cite{HR}, \cite{GHLS}, \cite{BK2}.  In this note we have argued that there is no fundamental obstruction to a well-defined 5D treatment of entropy functions with Chern-Simons terms, provided one implements the correct physical 5D charges  into the calculations. In general these 5D charges differ from those used in the dimensionally reduced approach due to spectral flow shifts.  However to fully specify a charge, one needs to obtain the equation of motion of the corresponding gauge field which is sourced by that charge. Within the setting of the entropy formalism, these gauge fields are determined via moduli $e^I$, $e^0$ and $a^I$.  Therefore upon extremising  ${\cal F}_5$  with respect to these moduli one can determine  the electric charges. On the other hand the magnetic charges are pre-fixed from the beginning. Our calculations demonstrate  that once the entropy function is expressed in terms of these physical 5D charges, it immediately falls into a 5D gauge invariant expression, even without requiring  to fix all the remaining moduli  $v_1$, $v_2$, $\omega$, $X^I$.  Moreover because of the fact that gauge fields and consequently charges of 5D geometries with different near-horizon topologies will in general be quite different, we find that one cannot construct a  universal entropy function that describes any 5D geometry in the presence of Chern-Simons terms and which is also gauge invariant.  In reference \cite{GJ}, they do manage to write down  a unified entropy function, however that can  only be expressed in terms of off-shell charges and  it is in fact not invariant under spectral flow transformations.  Therefore in order to check 5D gauge invariance,  we had to treat the $AdS_2 \times S^2 \times S^1$ black ring topology and the $AdS_2 \times S^3$  black hole topology separately. 

As is well-known, Chern-Simons terms in odd dimensions induce spectral flow shifts in the supergravity action, which also reflect in the defining notion of charges in these theories \cite{DM}.   In our analysis for the black ring, we have seen  that  these spectral flow equations also arise in a natural way out of Sen's formalism in 5D. Consequently the 5D electric charges were no longer gauge invariant and neither was the reduced action   ${\cal F}_5^{br}$.  Nonetheless the entropy function   ${\cal E}_5^{br}$ itself turned out to remain  invariant under gauge/spectral flow transformations  if it is expressed as a function of the correct physical charges. We have also verified  that the electric charges computed here from Sen's approach are identical to the Page charges expected from 5D supergravity : our charges calculations for the black ring give a precise match with the charge  integrals  recently computed  by  \cite{HOT}  on  the  basis   of  near-horizon data. 

On the other hand, whilst computing for the 5D black hole we found that the electric charges turned out not to be  Page but simply 5D Maxwell charges with no spectral flow shifts.  This was because a vanishing magnetic flux in an $AdS_2 \times S^3$ geometry suppresses all spectral flow shifts. As a consequence, the 5D charges of this black hole exactly match those of its 4D counterpart upon compactification of the fifth dimension. This corroborates with the 4D/5D lift  of \cite{GSY1}.  Within this set-up,  gauge invariance of the entropy function thereon  follows in a straightforward manner.  Then extremising   ${\cal E}_5^{bh}$   to  compute the black hole entropy indeed gave us an exact match with the result of \cite{L}, where the latter was obtained via an attractor mechanism calculation. This resolves the slight discrepancy in the result of \cite{GJ} where their entropy did not quite match \cite{L} : because their electric charges did not agree with  those of \cite{L}. Besides the comparison to \cite{L}, we have also provided additional evidence to support the claim that  $Q_I^{bh}$ computed  here are the correct charges to work with  by showing that they also match exactly with the  charges  of \cite{HOT}, which  were obtained from a 5D supergravity approach.  The discrepancy in the charges of \cite{GJ} arise whilst dimensionally reducing the Chern-Simons terms to 4D : namely, they assume an $x^{\mu}$-dependence for the moduli $a^I$; and only set the $a^I$  to constants in the final step. Consequently this introduces  terms in ${\cal L}_{CS}^{4D}$, which  incorrectly shift their electric charges, thereby causing a mismatch with the entropy of \cite{L}. However from the point of view of a manifestly  5D calculation, there was no natural way to assume such an  $x^{\mu}$-dependence  ( whilst already in the 5D near-horizon geometry ) and then abruptly deem them constants later in  the calculation. The 5D components of the field strength $F_{rt}$, $F_{\theta  \phi}$ are constants in the near-horizon geometry and giving the fields  an $x^{\mu}$-dependence through  $a^I$ would seem to come in conflict with the isometries of the near-horizon geometry. Moreover from the result of \cite{HOT} given in  eq.(\ref{2.11a}), the $ \int_{\Sigma}    6 C_{IJK} A^J \wedge F^K $ term vanishes for this black hole in the absence of an effective magnetic flux ( $ p^I + \tilde{a}^I  p^0$ ).  It is only the $ \int_{\Sigma}  \ast F_I $ term that  contributes to the charge.  Inserting the expression for the near-horizon field strength into the integral of eq.(\ref{2.11a}), exactly reproduces  our expression for  $Q_I^{bh}$.   The extra terms in the charges of \cite{GJ}  would simply not agree with the integral of \cite{HOT}.  This seems to suggest that assuming an $x^{\mu}$-dependence on any of the moduli in the  near-horizon geometry and then setting them to constants after dimensional reduction might be suspect. Within the entropy formalism, the isometries of the geometry are crucial to the analysis and all physical quantities ought to obey these. This imposes restrictions on the moduli, which works well when the latter are deemed  constants in this geometry at any stage of the analysis.  

A related line of interest which we have investigated in this note concerns black ring $\leftrightarrow$ black hole interpolation in the context of Sen's formalism.  The idea behind  such an  interpolation between geometries has been  familiar  since the work of \cite{EEMR3}, where it was shown using  black ring solutions from  \cite{EEMR1}, \cite{EEMR2}.  For what we had in mind here, it was more convenient to reformulate this interpolation using the most general 5D ${\cal N} = 1$ ungauged supergravity solution  of \cite{GGHPR},  \cite{GG3}  and varying the  Taub-NUT  modulus $l$ from a specified point to a vanishing limit. This way the structure of  harmonic functions and brane wrappings associated to the two geometries is more readily  manifest. The supergravity solution of course captures the global structure of the geometry, whereas the entropy formalism is only a near-horizon analysis. Therefore in principle it is not possible to construct a full-fledged interpolation of solutions using  the latter. However we have still managed to show within the Sen formalism that upon interchanging off-diagonal entries in the Kaluza-Klein metric bearing $e^0$ terms with those bearing $p^0$ ones, yields algrbraic data that can be compared to the limiting supergravity solutions in such a way that parameters in the  Kaluza-Klein metric can be specifically associated to  brane wrappings  in the supergravity solution for both the black ring and black hole. In retrospect, this also lends some physical intuition  to  the  ad hoc  assignment of variables made in the black hole/black ring  metric ansatz  proposed  in  \cite{GJ}.  Our original motivation in studying this $e^0 \leftrightarrow p^0$ exchange was in the hope of finding some sort of  black ring/black hole duality loosely speculated by  \cite{GJ}. However within the  context of our  analysis, the  $e^0 \leftrightarrow p^0$ exchange seems to relate more with the idea of a  geometric interpolation rather than  any  string or gravitational duality.  There is though an interesting work by \cite{BDF} which might be more in the direction of seeking such a string duality between 5D black holes and black rings. In that work, the authors propose a duality between  microstate degeneracies of  a D0-D2-D4 system with those of a D0-D2-D6 system on the same Calabi-Yau via a Fourier-Mukai transform.  From a 5D perspective, this would lift to a black hole/black string duality. From our discussion  in section 4, we have seen that the M-theory lift of a D2-D4-D6 system gives a 5D black hole, whereas a D0-D2-D4 system in the vicinity of a D6 charge, lifts to a black ring. It would therefore be quite interesting to  see if a microscopic duality along the lines of  \cite{BDF} can also be constructed for this black hole/black ring system.

\section*{Acknowledgements}
I would like to take this opportunity to thank Kevin Goldstein, Michele Maio, Erik Verlinde for useful discussions and especially Dumitru Astefanesei for extensive proof-reading of this manuscript. Financially of course, this research has been supported by De Stichting voor Fundamenteel Onderzoek der Materie (FOM); .....  but on an emotional and human level, by those closest to me.


\begin{thebibliography}{99}
\bibliographystyle{plain}



\bibitem{S1}
A.~Sen,
``Black hole entropy function and the attractor mechanism in higher  derivative gravity,''
JHEP {\bf 0509}, 038 (2005)
[arXiv:hep-th/0506177].


\bibitem{S2}
A.~Sen,
``Black Hole Entropy Function, Attractors and Precision Counting of Microstates,''
arXiv:0708.1270 [hep-th].


\bibitem{EEMR1}
H.~Elvang, R.~Emparan, D.~Mateos and H.~S.~Reall,
``A supersymmetric black ring,''
Phys.\ Rev.\ Lett.\  {\bf 93}, 211302 (2004)
[arXiv:hep-th/0407065].

\bibitem{EEMR2}
H.~Elvang, R.~Emparan, D.~Mateos and H.~S.~Reall,
``Supersymmetric black rings and three-charge supertubes,''
Phys.\ Rev.\  D {\bf 71}, 024033 (2005)
[arXiv:hep-th/0408120].


\bibitem{EEMR3}
H.~Elvang, R.~Emparan, D.~Mateos and H.~S.~Reall,
``Supersymmetric 4D rotating black holes from 5D black rings,''
JHEP {\bf 0508}, 042 (2005)
[arXiv:hep-th/0504125].


\bibitem{BK1}
I.~Bena and P.~Kraus,
``Three charge supertubes and black hole hair,''
Phys.\ Rev.\  D {\bf 70}, 046003 (2004)
[arXiv:hep-th/0402144].


\bibitem{B}
I.~Bena,
``Splitting hairs of the three charge black hole,''
Phys.\ Rev.\  D {\bf 70}, 105018 (2004)
[arXiv:hep-th/0404073].


\bibitem{BW1}
I.~Bena and N.~P.~Warner,
``One ring to rule them all ... and in the darkness bind them?,''
Adv.\ Theor.\ Math.\ Phys.\  {\bf 9}, 667 (2005)
[arXiv:hep-th/0408106].


\bibitem{BK3}
I.~Bena and P.~Kraus,
``Microscopic description of black rings in AdS/CFT,''
JHEP {\bf 0412}, 070 (2004)
[arXiv:hep-th/0408186].


\bibitem{BKW}
I.~Bena, P.~Kraus and N.~P.~Warner,
``Black rings in Taub-NUT,''
Phys.\ Rev.\  D {\bf 72}, 084019 (2005)
[arXiv:hep-th/0504142].


\bibitem{GJ}
K.~Goldstein and R.~P.~Jena,
``One entropy function to rule them all...''
arXiv:hep-th/0701221.


\bibitem{CP}
R.~G.~Cai and D.~W.~Pang,
``On Entropy Function for Supersymmetric Black Rings,''
JHEP {\bf 0704}, 027 (2007)
[arXiv:hep-th/0702040].


\bibitem{SS1}
B.~Sahoo and A.~Sen,
``BTZ black hole with Chern-Simons and higher derivative terms,''
JHEP {\bf 0607}, 008 (2006)
[arXiv:hep-th/0601228].


\bibitem{GSY1}
D.~Gaiotto, A.~Strominger and X.~Yin,
``New connections between 4D and 5D black holes,''
JHEP {\bf 0602}, 024 (2006)
[arXiv:hep-th/0503217].


\bibitem{GSY2}
D.~Gaiotto, A.~Strominger and X.~Yin,
``5D black rings and 4D black holes,''
JHEP {\bf 0602}, 023 (2006)
[arXiv:hep-th/0504126].


\bibitem{KL1}
P.~Kraus and F.~Larsen,
``Attractors and black rings,''
Phys.\ Rev.\  D {\bf 72}, 024010 (2005)
[arXiv:hep-th/0503219].


\bibitem{MS}
J.~F.~Morales and H.~Samtleben,
``Entropy function and attractors for AdS black holes,''
JHEP {\bf 0610}, 074 (2006)
[arXiv:hep-th/0608044].


\bibitem{L}
F.~Larsen,
``The attractor mechanism in five dimensions,''
arXiv:hep-th/0608191.


\bibitem{GGHPR}
J.~P.~Gauntlett, J.~B.~Gutowski, C.~M.~Hull, S.~Pakis and H.~S.~Reall,
``All supersymmetric solutions of minimal supergravity in five dimensions,''
Class.\ Quant.\ Grav.\  {\bf 20}, 4587 (2003)
[arXiv:hep-th/0209114].


\bibitem{GG3}
J.~P.~Gauntlett and J.~B.~Gutowski,
``All supersymmetric solutions of minimal gauged supergravity in five
dimensions,''
Phys.\ Rev.\ D {\bf 68}, 105009 (2003)
[Erratum-ibid.\ D {\bf 70}, 089901 (2004)]
[arXiv:hep-th/0304064].


\bibitem{KK}
R.~Kallosh and B.~Kol,
``E(7) Symmetric Area of the Black Hole Horizon,''
Phys.\ Rev.\  D {\bf 53}, 5344 (1996)
[arXiv:hep-th/9602014].


\bibitem{BMPV}
J.~C.~Breckenridge, R.~C.~Myers, A.~W.~Peet and C.~Vafa,
``D-branes and spinning black holes,''
Phys.\ Lett.\  B {\bf 391}, 93 (1997)
[arXiv:hep-th/9602065].


\bibitem{GMT}
J.~P.~Gauntlett, R.~C.~Myers and P.~K.~Townsend,
``Black holes of D = 5 supergravity,''
Class.\ Quant.\ Grav.\  {\bf 16}, 1 (1999)
[arXiv:hep-th/9810204].


\bibitem{MSW}
J.~M.~Maldacena, A.~Strominger and E.~Witten,
``Black hole entropy in M-theory,''
JHEP {\bf 9712}, 002 (1997)
[arXiv:hep-th/9711053].


\bibitem{CGMS}
M.~Cyrier, M.~Guica, D.~Mateos and A.~Strominger,
``Microscopic entropy of the black ring,''
Phys.\ Rev.\ Lett.\  {\bf 94}, 191601 (2005)
[arXiv:hep-th/0411187].


\bibitem{HR}
G.~T.~Horowitz and H.~S.~Reall,
``How hairy can a black ring be?,''
Class.\ Quant.\ Grav.\  {\bf 22}, 1289 (2005)
[arXiv:hep-th/0411268].


\bibitem{GHLS}
M.~Guica, L.~Huang, W.~W.~Li and A.~Strominger,
``R**2 corrections for 5D black holes and rings,''
JHEP {\bf 0610}, 036 (2006)
[arXiv:hep-th/0505188].


\bibitem{BK2}
I.~Bena and P.~Kraus,
``R**2 corrections to black ring entropy,''
arXiv:hep-th/0506015.


\bibitem{OSV}
H.~Ooguri, A.~Strominger and C.~Vafa,
``Black hole attractors and the topological string,''
Phys.\ Rev.\  D {\bf 70}, 106007 (2004)
[arXiv:hep-th/0405146].


\bibitem{BCDMV}
J.~de Boer, M.~C.~N.~Cheng, R.~Dijkgraaf, J.~Manschot and E.~Verlinde,
``A farey tail for attractor black holes,''
JHEP {\bf 0611}, 024 (2006)
[arXiv:hep-th/0608059].


\bibitem{KL2}
P.~Kraus and F.~Larsen,
``Partition functions and elliptic genera from supergravity,''
JHEP {\bf 0701}, 002 (2007)
[arXiv:hep-th/0607138].


\bibitem{DM}
D.~Marolf,
``Chern-Simons terms and the three notions of charge,''
arXiv:hep-th/0006117.


\bibitem{AGJST}
D.~Astefanesei, K.~Goldstein, R.~P.~Jena, A.~Sen and S.~P.~Trivedi,
``Rotating attractors,''
JHEP {\bf 0610}, 058 (2006)
[arXiv:hep-th/0606244].


\bibitem{HOT}
K.~Hanaki, K.~Ohashi and Y.~Tachikawa,
``Comments on charges and near-horizon data of black rings,''
JHEP {\bf 0712}, 057 (2007)
[arXiv:0704.1819 [hep-th]].


\bibitem{KLR}
H.~K.~Kunduri, J.~Lucietti and H.~S.~Reall,
``Near-horizon symmetries of extremal black holes,''
Class.\ Quant.\ Grav.\  {\bf 24}, 4169 (2007)
[arXiv:0705.4214 [hep-th]].


\bibitem{AY}
D.~Astefanesei and H.~Yavartanoo,
``Stationary black holes and attractor mechanism,''
Nucl.\ Phys.\  B {\bf 794}, 13 (2008)
[arXiv:0706.1847 [hep-th]].


\bibitem{XDA}
X.~D.~Arsiwalla,
``More Rings to rule them all : Fragmentation, 4D/5D and Split-Spectral Flows,''  JHEP {\bf 0802}, 066 (2008)
[arXiv:0709.0308 [hep-th]].


\bibitem{BDF}
I.~Bena, D.~E.~Diaconescu and B.~Florea,
``Black string entropy and Fourier-Mukai transform,''
JHEP {\bf 0704}, 045 (2007)
[arXiv:hep-th/0610068].




\end{thebibliography}
\end{document}